# Infrared polarizer based on direct coupling to surface-plasmon polaritons

Alireza Shahsafi[1], Jad Salman[1], Bryan E. Rubio Perez[1], Yuzhe Xiao[1], Chenghao Wan[1,2], and Mikhail A. Kats[1,2,3]

[1]Department of Electrical and Computer Engineering, University of Wisconsin-Madison
[2]Department of Materials Science and Engineering, University of Wisconsin-Madison
[3]Department of Physics, University of Wisconsin-Madison

**Abstract:** We propose a new type of reflective polarizer based on polarization-dependent coupling to surface-plasmon polaritons (SPPs) from free space. This inexpensive polarizer is relatively narrowband but features an extinction ratio of up to 1000 with efficiency of up to 95% for the desired polarization (numbers from a calculation), and thus can be stacked to achieve extinction ratios of $10^6$ or more. As a proof of concept, we experimentally realized a polarizer based on nanoporous aluminum oxide that operates around a wavelength of 10.6 μm, corresponding to the output of a $CO_2$ laser, using aluminum anodization, a low-cost electrochemical process.

Optical polarizers typically function in transmission, reflection, or refraction modes. Wire-grid polarizers based on subwavelength parallel metallic wires transmit light with polarization perpendicular to the wire direction [1], [2]. Glan–Taylor prisms are refractive polarizers [3] that spatially segregate light with orthogonal linear polarizations. Reflective polarizers reflect light with a particular polarization; for example, Brewster-angle polarizers selectively reflect light polarized out of the plane of incidence (s polarization) [4].

The most common high-extinction-ratio polarizers in the long-wavelength infrared (LWIR) are based on finely spaced wire grids or other micro/nanostructures and are thus rather expensive optical components. Here, we introduce an infrared reflective-type polarizer based on direct coupling between free-space light and a special type of surface-plasmon polariton (SPP) mode that has a dispersion relation above the light line. This approach requires no micro- or nanopatterning and can thus be inexpensively manufactured. We demonstrated such a polarizer based on porous anodized aluminum, designed to operate at $CO_2$-laser wavelengths.

SPPs are surface waves that propagate at the interface between a metal and a dielectric [5]. To excite SPPs from free space, incident light with a certain frequency and wave vector (projected onto the direction of the interface) are required. Usually, the wave vector of the incident light has a smaller value than the SPP wave vector at the same frequency, making coupling difficult. A standard approach to enable coupling uses prisms comprising high-index dielectric materials to increase the wave vector of the incident light to match with that of the SPP—in particular, in the Kretschmann [6] [Fig. 1(a)] and Otto [7] [Fig. 1(b)] configurations. In the Kretschmann configuration, light inside the prism reaches a thin metal layer and tunnels through to the SPP mode on the other side [Fig. 1(a)]. In the Otto configuration, light tunnels through a small dielectric gap before reaching the SPP mode [Fig. 1(b)]. Note that, in either case, SPPs can only be excited by light polarized parallel to the plane of incidence (p polarization), whereas light in the other polarization is mostly reflected [5].

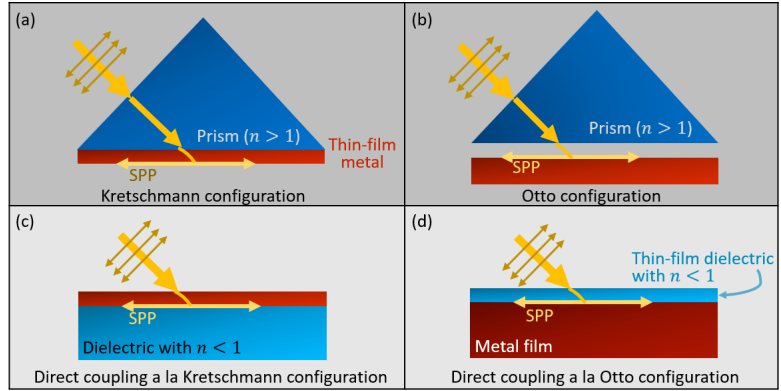

**Figure 1. (a, b)** Excitation of SPPs using a prism in the (a) Kretschmann configuration [6] and (b) Otto configuration [7]. **(c, d)** Prism-free direct coupling to SPPs in the modified (a) Kretschmann and (b) Otto configurations by using a dielectric with real part of its refractive index less than one, resulting in a decrease of the SPP wavevector [8].

A complementary approach can also be taken to excite SPPs [Fig. 1(c)]: the wave vector of SPPs can be decreased by replacing a typical dielectric at the metal/dielectric interface with a dielectric that has real part of its refractive index lower than unity ($n < 1$). We have previously demonstrated this type of SPP coupling in a Kretschmann-like configuration using a 10-nm gold layer deposited onto an $SiO_2$ wafer in the wavelength range of 7.4 to 7.7 µm, where the real part of the $SiO_2$ refractive index is around 0.8 due to the presence of vibrational resonances at a slightly longer wavelength [8]. Here, we take a similar approach, but in the Otto configuration (without the prism) [Fig. 1(d)], to create our reflective polarizer. We note that coupling to such a leaky SPP mode is sometimes referred to as the Berreman effect [9]–[11], though we find the connection to the well-known Otto and Kreschmann configurations for SPP coupling to be more instructive and thus prefer that language.

Around a wavelength of 10 µm, amorphous $Al_2O_3$ has $n < 1$ and modest optical losses, represented by a small extinction coefficient, $\kappa$, as shown in Fig. 2(a-b) [12]. Using these optical properties and the refractive index of aluminum, we calculated angle- and polarization-dependent reflectance for a 1-µm-thick layer of amorphous $Al_2O_3$ on an optically thick aluminum substrate using the transfer-matrix method. The thickness of the $Al_2O_3$ layer is approximately one tenth of the target free-space wavelength for our polarizer (free-space wavelength of 10.6 µm). Figure 2(c) shows our calculation of the angle-dependent reflectance of this $Al_2O_3$/Al structure for both p- and s-polarized light ($R_p$ and $R_s$). $R_p$ approaches zero around an incident angle of 65° for a wavelength of 10.2 µm due to coupling to SPPs, while $R_s$ remains close to unity for all wavelengths and angles.

Polarizers are readily evaluated using two figures of merit: extinction ratio and efficiency. The extinction ratio is the ratio of the output power to the input power, given input light polarized perpendicular to the desired polarization (i.e., in the "wrong" polarization). The efficiency is the ratio of the output power to the input power, given input light polarized along the desired polarization direction. Therefore, in Fig. 2(c), $R_s$ can be interpreted as the efficiency of a reflective polarizer based on direct coupling to SPPs, and thus we observe that such a polarizer can have efficiency ($R_s$) above 90%, with an extinction ratio ($R_s/R_p$) approaching infinity as $R_p$ approaches zero.

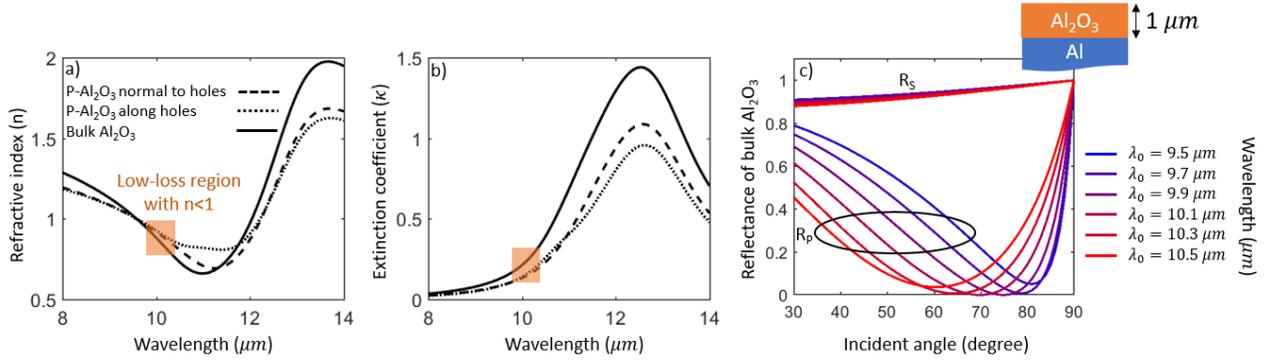

**Figure 2. (a)** Real ($n$) and **(b)** imaginary ($\kappa$) parts of the complex refractive index of bulk (non-porous) amorphous Al$_2$O$_3$ (solid) [12] and porous anodic aluminum oxide (p-AAO), for polarization along (dotted) and perpendicular (dashed) to the p-AAO holes. The p-AAO calculation is based on Maxwell-Garnett effective-medium theory [13]. **(c)** P- and s-polarized calculated reflectance for a 1-µm-thick layer of amorphous Al$_2$O$_3$ on top of a thick aluminum film for the wavelength range in which Al$_2$O$_3$ has refractive index less than one and moderate loss.

Since our target wavelength was 10.6 µm rather than the ideal operating wavelength of 10.2 µm for the structure in Fig. 2(c), we performed several calculations to identify an appropriate geometry. Our primary degrees of freedom were the Al$_2$O$_3$ thickness and porosity, though we also had some flexibility with the incident angle. In particular, increasing the porosity (density of air inclusions) of the Al$_2$O$_3$ is expected to bring the refractive index closer to that of air and reducing optical loss, resulting in coupling to SPPs at a slightly longer wavelength compared to the simulations in Fig. 2(c). Based on our calculations, we identified a porosity of approximately 0.35 as a candidate for a 10.6-µm polarizer (see Supporting Info 1).

To realize our structure without the use of expensive nanofabrication techniques, we relied on an electrochemical anodization process to grow porous anodic aluminum oxide (p-AAO) with porosity of 0.35 on top of aluminum sheets (synthesis by InRedox LLC). Anodization can grow oxide layers on the surfaces of certain metals (aluminum, titanium, etc.), and is often used to enhance surface durability or for decorative purposes [14].

Figure 3 shows scanning electron microscope (SEM) images of the resulting p-AAO. Similarly to p-AAO in the literature [14], the pores are cylindrical all the way through the oxide film. Here, the diameter of the pores is approximately 100 nm, which is much smaller than our working wavelength (see Supporting Info 3 for more details). Based on the SEM images, we modeled the optical properties of p-AAO based on an effective-medium theory for anisotropic media [13]:

$$n_\| = \sqrt{(1-p) \times n_{AO}^2 + p \times n_{air}^2}, \qquad \text{Eq. 1}$$

$$n_\perp = n_{AO}\sqrt{\frac{2(1-p) \times n_{AO}^2 + (1+2p) \times n_{air}^2}{(2+p) \times n_{AO}^2 + (1-p) \times n_{air}^2}}. \qquad \text{Eq. 2}$$

Here, $p$ is the volume ratio of air and $n_{AO}$ is the complex refractive index of bulk Al$_2$O$_3$. The real and imaginary parts of the anisotropic complex index for $p = 0.35$ are shown in Fig. 1 (a-b). We used these optical properties to model the reflectance of our polarizers. The calculated spectral reflectance for three different incidence angles for 1-µm-thick p-AAO with porosity of 0.35 on an aluminum substrate are shown in Fig. 3(b), indicating an efficiency of more than 90%.

We performed angle-dependent spectral measurements of the fabricated p-AAO/Al structures using the reflection mode of the J. A. Woollam IR-VASE Mark II ellipsometer, and found excellent agreement with the calculations [Fig. 3(b)]. The reduction of the measured $R_s$ value (~0.85) compared to the calculations (0.95) is assumed to be mostly due to scattering from inhomogeneities in the structure. Calculations describing SPP polarizers in different wavelength ranges based on different dielectric materials are provided in Supporting Info, Section 3.

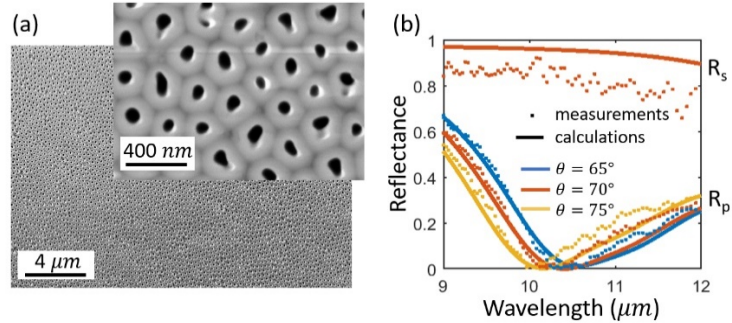

**Figure 3. (a)** Scanning electron microscope (SEM) images of porous anodic aluminum oxide (p-AAO), imaged from the top of the sample. The inset SEM image of the same sample magnified 10 times. **(b)** Measured (dotted) and calculated (solid) spectral reflectance for p polarization, $R_p$, and s polarization, $R_s$, of a 1-µm layer of p-AAO on top of a thick aluminum film for the wavelength range in which $Al_2O_3$ has refractive index less than one. Different colors show incident light with different incident angles.

The extinction ratio and operational bandwidth of our proposed polarizer can be increased by having two or more p-AAO/Al samples positioned next to each other in a way that the reflected light from the first one is incident on the second one, etc. [e.g., Fig. 4(a)], though this decreases the efficiency of the resulting polarizer. We used a simple scheme called the K-geometry [15] which has a mirror between two samples to guide the reflected light from the first sample to the next (the mirror can also be replaced with another p-AAO/Al sample). This scheme results in a polarizer that maintains the position and directionality of an incident beam, so it can be incorporated into the path of a laser beam without affecting any other optical components [Fig. 4(b)]. Note that by slightly rotating the orientation of the individual films, the total operational bandwidth of the polarizer can be increased, albeit with a reduction in the extinction ratio.

To test the operation of the proposed polarizer, we 3D printed a box [Fig. 4(b)] that holds two p-AAO/Al samples like those in Fig. 3, designed to function as polarizers at an incident angle of 70°, for wavelengths corresponding to $CO_2$ laser emission (10.2 to 10.7 µm) [16]. We measured the reflected light using only one p-AAO/Al sample in the box for p- and s-polarized light [Fig. 4(c-d)], using two metallic mirrors to guide the light without changing the direction or position of the beam from the input to the output. The corresponding wavelength-dependent extinction ratio ($R_p/R_s$), both theoretical and measured using a $CO_2$ laser with power of approximately 200 mW, is shown in Fig. 4(e).

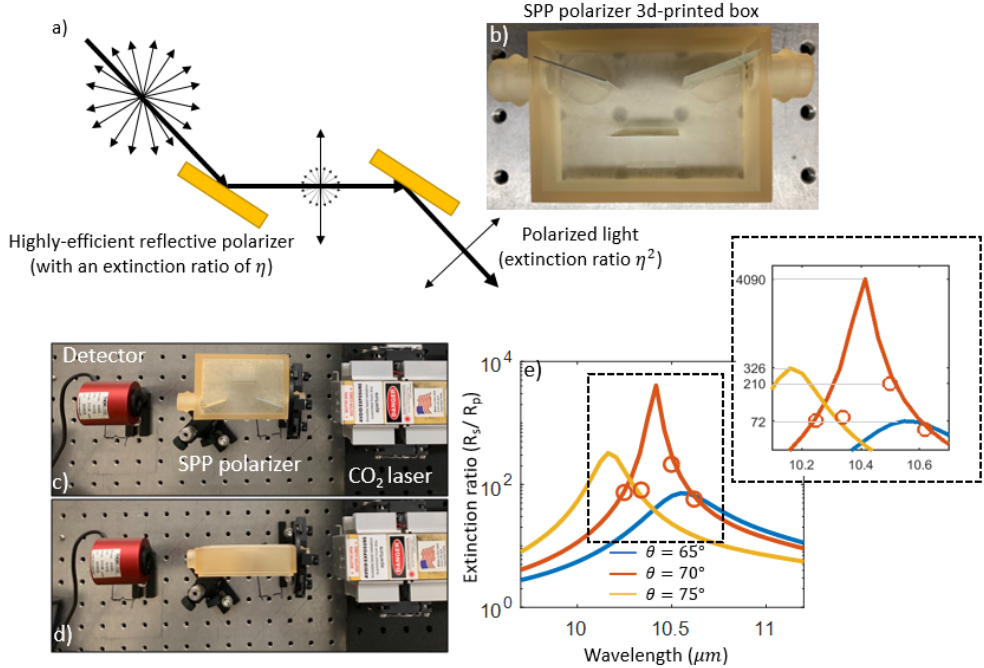

**Figure 4.** Increasing the extinction ratio or operational bandwidth of our SPP polarizer by placing two samples next to each other **(a)** without using a mirror, **(b)** by using a mirror, resulting in conservation of beam position and direction. **(c-d)** Testing of the polarizer box shown in (b), but with a single p-AAO/Al sample and two metallic mirrors (i.e., here we are not stacking the p-AAO/Al polarizers). Measurements were performed with a $CO_2$ laser for two different orientations of the polarizer. **(e)** Calculated (lines) and experimental (circles) extinction ratios ($R_s/R_p$,) using the setup in c, d.

## Conclusion

We proposed a reflective polarizer based on direct coupling to surface plasmon polaritons (SPPs), enabled by the use of a dielectric with a refractive index less than unity. We demonstrated this concept using porous anodized aluminum oxide (p-AAO), making a polarizer optimized for the $CO_2$-laser wavelength. Notably, our method does not require any nanopatterning or other expensive fabrication techniques, resulting in a low-cost design that can be easily mass produced.

## Acknowledgement

We acknowledge funding from the Office of Naval Research (N00014-20-1-2297). Various characterization was performed at the Nanoscale Imaging and Analysis Center and the Soft Materials Characterization Laboratory, core facilities at UW-Madison. We acknowledge use of facilities and instrumentation supported by NSF through the University of Wisconsin Materials Research Science and Engineering Center (DMR-1720415). We thank Orad Reshef for identifying refs. [9]–[11] at the preprint stage.

Supporting information for

# Infrared polarizer based on direct coupling to surface-plasmon polaritons

Alireza Shahsafi[1], Jad Salman[1], Bryan E. Rubio Perez[1], Yuzhe Xiao[1], Chenghao Wan[1,2], and Mikhail A. Kats[1,2,3]

[1]Department of Electrical and Computer Engineering, University of Wisconsin-Madison
[2]Department of Materials Science and Engineering, University of Wisconsin-Madison
[3]Department of Physics, University of Wisconsin-Madison

## S1. Anodization process and coupling to SPPs in porous samples

We used anodization, an electrochemical process, to synthesize our porous aluminum oxide (p-AAO) on aluminum sheets. The process converts metallic surfaces into oxide layers resulting in more durable and corrosion-resistant surfaces [S1]. The resulting oxide layer becomes uniformly porous at certain growth conditions (chemistry, current, temperature, etc.). We found that a p-AAO layer with porosity of 0.35 and thickness of 1 micron can be a good candidate for our polarizer. This finding is a result of various thin-film calculations and measurements using many different samples with different porosities purchased from InRedox LLC to validate our modeling.

In general, increasing the porosity results in a red shift of the coupling wavelength according to our calculation based on the anisotropic refractive index of p-AAO [S2]. This is shown in the Fig. S1, where larger porosity results in coupling at a longer wavelength. This happens because when the refractive index is less than one, introducing air voids shifts the refractive index to closer to unity.

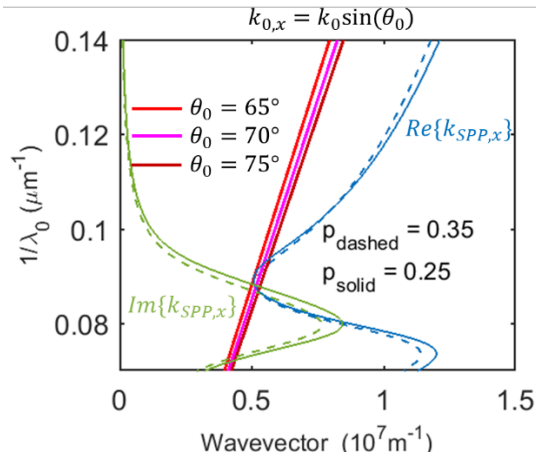

**Figure S1**. Relation between the wave vector and the reciprocal of the free-space wavelength (i.e., the angular frequency) for surface-plasmon polaritons (SPPs) at the interface between Al and p-AAO for different porosity (dashed and solid lines are representative of 0.35 and 0.25 porosity, respectively). Three light lines show the $k_{0,x}$ component of the wave vector of the incident light for three incident angles. The blue lines show the real part of $k_{SPP,x}$ and green lines show the imaginary part of $k_{SPP,x}$.

Figure S2 shows the wavelength and thickness dependence of the extinction ratio of the p-AAO/Al structure for different incident angles and porosities. Based on this figure, very high extinction ratios ( > 1000) can be achieved. Increasing the porosity tends to push the region with high extinction ratios toward longer wavelengths and larger thicknesses. Increasing the incident angle tends to increase the extinction ratio.

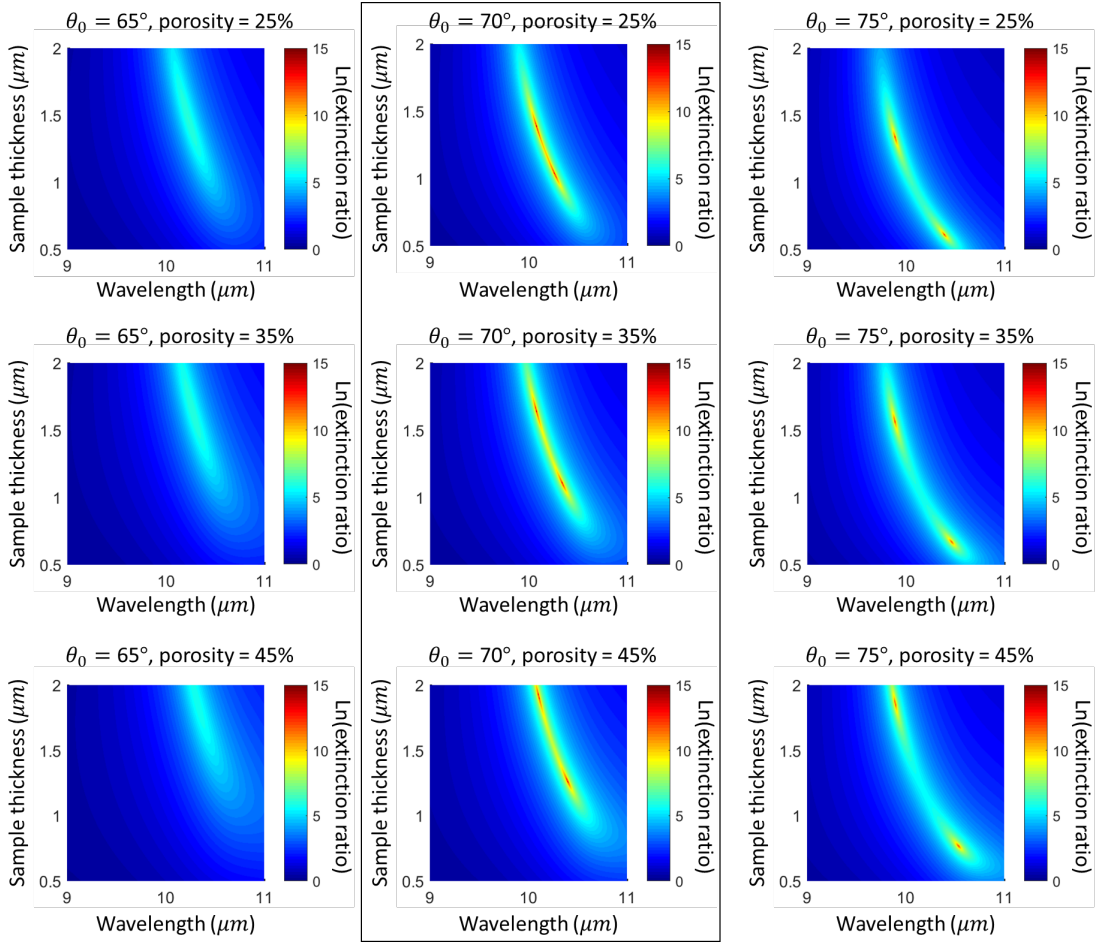

Figure S2) Calculated wavelength and thickness dependence of the extinction ratio of a p-AAO/Al sample for different incident angles and porosities.

Furthermore, we conducted a calculation for a finer angular step size (from 65° to 75°, with step size of 1°) to investigate the difference between the calculations and our measurement using a $CO_2$ laser (Fig. S3). This calculation shows that the peak of extinction ratio has a large angular sensitivity. Thus, it is possible that some of the differences between the calculations and our measurements are a result of errors in the incident angle on the sample due to misalignment.

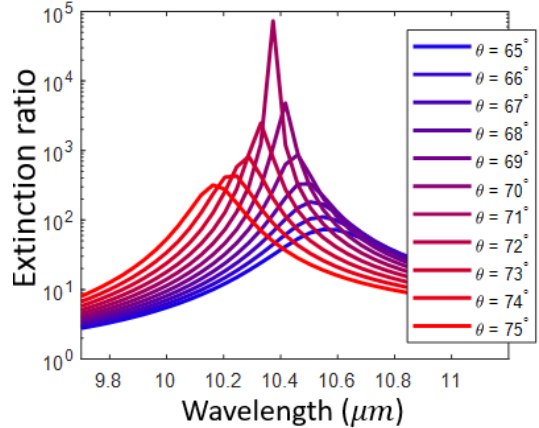

Figure S3) Calculated extinction ratio of our p-AAO/Al sample versus wavelength for incident angles from 65° to 75° with step size of 1°.

## S2. Using different dielectrics: covering other spectral ranges

To design SPP polarizers at substantially different wavelengths, different materials with refractive index less than unity must be used. Fig. S4(a) shows different candidate dielectric materials with refractive index less than one. In particular, we are plotting a figure of merit (FOM) that we defined as $\kappa/(1-n)$ [S3]. Calculated results using two different dielectric materials with refractive index less than unity at two wavelength ranges, all on top of aluminum substrates, are shown in Fig. S4 (b-e). In Fig. S4(b, c), we show that indium tin oxide (ITO) can realize polarizers in the near infrared. In Fig. S4 (d, e), we show that calcium fluoride ($CaF_2$) can be used for far-infrared polarizers.

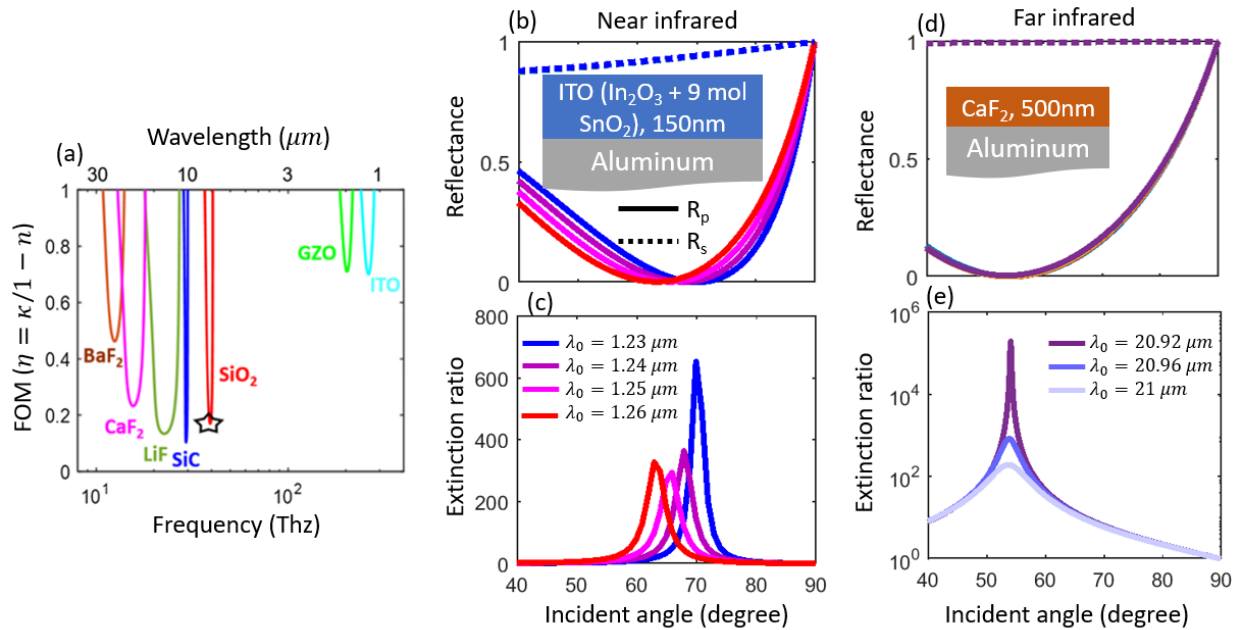

**Figure S4.** (a) Selection of other candidate dielectric materials with $n < 1$, including transparent conducting oxides in the near-infrared region and dielectric materials with strong vibrational resonances in the mid-infrared region (material data from [S4-5]). In all cases, the substrate is optically thick aluminum. (b) Reflectance and (c) extinction ratio of a near-infrared SPP polarizer based on a 150-nm film of indium tin oxide (ITO), with the optical properties from [S4]. (d) Reflectance and (e) extinction ratio of a far-infrared SPP polarizer based on a 500-nm CaF$_2$ thin film [S5].

### S3. Analysis of aluminum oxide surface

We analyzed the porosity of the samples based on the dark areas in the SEM images in Fig. S5. We assumed that these dark pixels in each SEM image correspond to the pores, and based on this assumption, we calculated surface porosities and diameters of the pores. For these images, this calculation yields diameters of 98 nm and 90 nm, respectively.

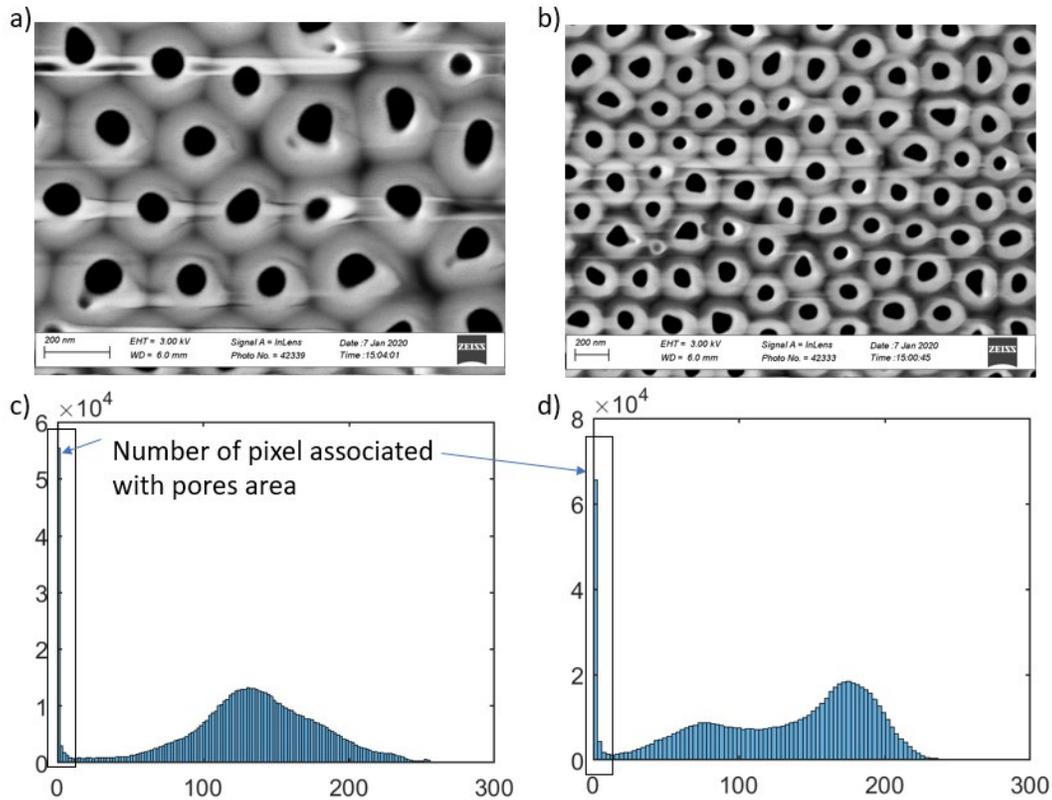

Figure S5) (a-b) Scanning electron microscope (SEM) images of the porous aluminum oxide surface and (c-d) histogram of the luminance of the SEM images.

Furthermore, we calculated the porosity across a large-area SEM in Fig. S6, and found there the porosity varies from about 0.31 to 0.44 across the image. We calculated the corresponding extinction ratios at incident angle of 75° versus wavelength for range of porosity (Fig. S6). Depending on the variation in the porosity, the extinction ratio can vary substantially.

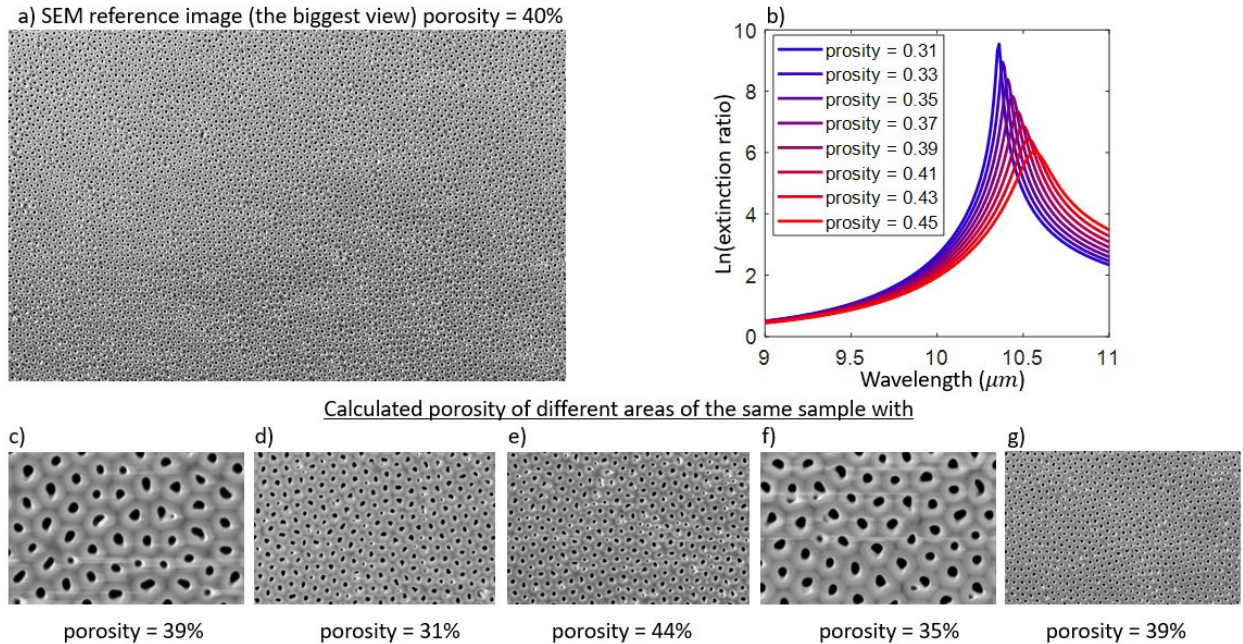

Figure S6) (a) Large-area SEM image to investigate the uniformity of the porosity of the samples. By analyzing the sub-regions of (a) in (c-g), we found that the porosity can vary from ~30 to ~45%. (b) Calculations of the extinction ratio corresponding to the porosities observed in c-g.

**Supplementary references**

[S1] W. Lee and S. Park, "Porous Anodic Aluminum Oxide : Anodization and Templated Synthesis of Functional Nanostructures," 2014

[S2] G. A. Niklasson and C. G. Granqvist, "Effective medium models for the optical properties of inhomogeneous materials," Appl. Opt., vol. 20, no. 1, p. 26, 1981.

[S3] A. Shahsafi et al., "Mid-infrared Optics Using Dielectrics with Refractive Indices below Unity," Phys. Rev. Appl., vol. 10, no. 3, p. 1, 2018.

[S4] J. Kim, G. V Naik, N. K. Emani, U. Guler, and A. Boltasseva, "Plasmonic Resonances in Nanostructured Transparent Conducting Oxide Films," IEEE J. Sel. Top. Quantum Electron., vol. 19, no. 3, p. 4601907, 2013.

[S5] W. Kaiser, W. G. Spitzer, R. H. Kaiser, L. E. Howarth, "Infrared Properties of CaF2, SrF2, and BaF2," Phys. Rev., vol. 127, no. 6, p. 1950, 1962.